\documentclass[conference]{IEEEtran}

\usepackage{cite}
\usepackage{amsmath,amssymb,amsfonts}

\usepackage{algorithm} 
\usepackage{algorithmic} 

\usepackage{graphicx}
\usepackage{multirow}
\usepackage{subcaption}
\usepackage{caption}
\usepackage{float}
\usepackage{textcomp}
\usepackage{xcolor}
\usepackage{siunitx}
\usepackage{tikz}
\usepackage{pgfplots}
\usepackage{stfloats}
\usepackage{mathrsfs}
\usepackage{balance}
\usepackage{amsthm}

\usepackage{comment}
\usepackage{booktabs}
\usepackage{acronym}
\usepackage[hyperfirst=true]{glossaries}
\usepackage[hidelinks]{hyperref} 
\usepackage{makecell}
\makeglossaries
\newacronym{ECG}{ECG}{electrocardiogram}
\newacronym{EMG}{EMG}{electromyogram}
\newacronym{DL}{DL}{deep learning}
\newacronym{EMD}{EMD}{empirical mode decomposition}
\newacronym{IMF}{IMF}{intrinsic mode function}
\newacronym{EMD-Conv}{EMD-Conv}{EMD-conventional}
\newacronym{EMD-IT}{EMD-IT}{EMD-interval thresholding}
\newacronym{EMD-ITF}{EMD-ITF}{EMD-improved thresholding function}
\newacronym{EMD-Custom}{EMD-Custom}{EMD-customized thresholding}
\newacronym{DWT}{DWT}{discrete wavelet transform}
\newacronym{QMF}{QMF}{quadrature mirror filter}
\newacronym{SNR}{SNR}{signal-to-noise ratio}
\newacronym{PPSNR}{PPSNR}{peak-to-peak signal-to-noise ratio}
\newacronym{RMSE}{RMSE}{root mean squared error}
\newacronym{PCC}{PCC}{Pearson correlation coefficient}
\newacronym{DAE}{DAE}{denoising autoencoder}
\newacronym{SDAE}{SDAE}{stacked denoising autoencoder}
\newacronym{PINN}{PINN}{physics-informed
neural network}
\newacronym{WBAN}{WBAN}{wireless body-area network}
\newacronym{WSN}{WSN}{wireless sensor network}
\newacronym{MSE}{MSE}{mean squared error}
\usepackage{orcidlink}
\usepackage{amsmath,amssymb,amsgen,amsbsy,amsopn,amsthm,bm,calc,epsfig,multicol,multirow,graphicx,psfrag}
\usepackage{tikz} 
\usetikzlibrary{shapes.callouts, positioning, calc, arrows.meta}
\newcommand{\T}{\mathsf{T}}
\usepackage{tikz}
\usepackage{tikz-3dplot}
\usetikzlibrary{arrows.meta,calc,backgrounds,patterns,
decorations.pathmorphing,
decorations.pathreplacing,
decorations.markings,positioning,shapes,
shapes.geometric, arrows} 
\usetikzlibrary{positioning, fit, arrows.meta}
\usepackage{pgfplots}
\usepgfplotslibrary{groupplots} 
\pgfplotsset{compat=1.18}       
\usepackage{stfloats}
\usepackage{mathrsfs}
\usepackage{multirow} 
\usepackage{hyperref}
\hypersetup{
    colorlinks=true,   
    linkcolor=blue, 
    citecolor=blue,  
    urlcolor=black 
}

\allowdisplaybreaks

\begin{document}

\title{A Comparative Study of ECG Denoising Methods for Wearable Applications}

\author{\IEEEauthorblockN{Bamrung Tausiesakul\IEEEauthorrefmark{2}, Anna Marcucci\IEEEauthorrefmark{1}, Amin Damrah\IEEEauthorrefmark{1}, Mauro Marchese\IEEEauthorrefmark{1}, Pietro Savazzi\IEEEauthorrefmark{1}\IEEEauthorrefmark{3}, Anna Vizziello\IEEEauthorrefmark{1}\IEEEauthorrefmark{3}} 
\IEEEauthorblockA{\IEEEauthorrefmark{1}Department of Electrical, Computer and Biomedical Engineering, University of Pavia, Italy\\ \IEEEauthorrefmark{2}Department of Electrical Engineering, Faculty of Engineering, Mahidol University, Nakhon Pathom, Thailand \\ \IEEEauthorrefmark{3}CNIT Consorzio Nazionale Interuniversitario per le Telecomunicazioni, Pavia, Italy \\
E-mail: mauro.marchese01@universitadipavia.it}
}

\maketitle

\begin{abstract}
Reliable \gls{ECG} monitoring in wearable and space environments
requires effective denoising of signals corrupted by non-stationary
\gls{EMG} interference. This paper presents a comparative
evaluation of model-based and DL-based denoising techniques for
upper-arm \gls{ECG} recordings acquired under real conditions. The model-based
methods include three \gls{EMD} variants and a
\gls{DWT} approach, while the \gls{DL} side is
represented by a \gls{SDAE} and a \gls{PINN}. All methods are evaluated on real
acquisitions under both relaxed and voluntary muscle contraction conditions,
using \gls{RMSE}, Pearson correlation, and \gls{PPSNR} as performance metrics.
Results reveal a fundamental trade-off: DL methods achieve superior
morphological reconstruction, while \gls{DWT} provides the strongest noise
suppression, highlighting complementary strengths for wearable cardiac
monitoring applications.
\end{abstract}

\begin{IEEEkeywords}
ECG, denoising, deep learning.
\end{IEEEkeywords}

\glsresetall 
\section{Introduction}

Continuous and reliable monitoring of cardiac activity is a cornerstone of both
clinical medicine and remote physiological surveillance. The \gls{ECG} remains the primary non-invasive tool for the assessment of cardiovascular
function, enabling real-time detection of arrhythmias, ischemic events, and
conduction abnormalities~\cite{serhani2020ecg}. Beyond traditional hospital
settings, the demand for unobtrusive long-duration cardiac monitoring has grown
substantially, driven by applications ranging from elderly care and post-surgical
follow-up to human health monitoring in extreme environments such as outer space \cite{10789688}.

In space missions, continuous physiological monitoring of astronauts is a
mission-critical requirement \cite{4649272}. The human cardiovascular system undergoes profound
adaptations in microgravity---including cardiac remodeling, fluid shifts, and
increased arrhythmia susceptibility---making real-time ECG surveillance
indispensable \cite{6091059}. These operational constraints call for compact, wearable, and
energy-efficient sensing platforms, precisely the kind of systems addressed by \glspl{WBAN} and \glspl{WSN} \cite{hayajneh2014survey, cheikhrouhou2016secure}. In
such resource-constrained deployments, signal quality is paramount: any
noise-induced degradation directly compromises the reliability of downstream
cardiac analysis and crew safety.

Conventional chest-mounted \gls{ECG} systems are poorly suited to prolonged wearable
use, motivating the exploration of alternative electrode configurations~\cite{francisco2021cardiac}.
Upper-arm \gls{ECG} acquisition has recently emerged as a promising wearable
alternative~\cite{Marcucci2025}, compatible with both \gls{WSN} nodes and space-suit
integration. However, the low amplitude of upper-arm \gls{ECG} recordings makes them
highly susceptible to \gls{EMG} interference from surrounding
muscles, which can severely distort the QRS complex and other clinically relevant
features, posing a fundamental challenge to automated \gls{ECG} analysis \cite{lazaro2020wearable}. Effective denoising is therefore a critical pre-processing stage in any wearable
\gls{ECG} pipeline.

 Among model-based approaches,  \gls{EMD} 
has attracted considerable interest for its adaptive data-driven decomposition
into \glspl{IMF}, well-suited to non-stationary biomedical
signals,  and has led to several denoising variants based on IMF selection and thresholding strategies~\cite{Mohguen2017EMDITF}.
The \gls{DWT} has likewise
established itself as a reference tool for \gls{ECG} denoising through its
multiresolution time-frequency representation,
studied across different mother wavelets,
decomposition
levels,
and threshold rules~\cite{mallat1989theory}.
On the \gls{DL} side, \glspl{SDAE} have
demonstrated robust feature learning under noise corruption~\cite{2016:Xiong_et_al},
while \glspl{PINN} \cite{2025:Zhu_et_al} have been proposed
to enforce temporal smoothness by embedding cardiac kinematics into the training
objective.

Despite this substantial body of work, most existing studies rely on synthetically
corrupted datasets that may not reflect the non-stationary, muscle-dependent
interference of real wearable acquisitions. This paper addresses this gap through
a systematic comparison of model-based and \gls{DL}-based methods on real upper-arm \gls{ECG}
data acquired under relaxed and voluntary muscle contraction conditions, providing
a realistic assessment relevant to space and \gls{WSN} deployments.


\section{System Model}
\label{sec:system_model}

To assess the effectiveness of the denoising methods, the upper-arm \gls{ECG} denoising problem is formulated as the estimation of an underlying clean cardiac signal from a noisy observation.

Let $x[n]$ denote the underlying clean upper-arm \gls{ECG} signal, where $n \in \{1,2,\dots,N\}$ is the discrete-time sample index and $N$ is the total number of samples within the observation window. The actual upper-arm \gls{ECG} acquisition, denoted as $y[n]$, can be modeled as

\begin{equation}
    y[n] = x[n] + v[n]\Longleftrightarrow \bm{y}=\bm{x}+\bm{v}\in\mathbb{R}^{N\times 1},
\end{equation}

where $v[n]$ represents the aggregate interference affecting the acquisition. 
In upper-arm \gls{ECG} monitoring, the dominant disturbance is typically generated by muscular activity and \gls{EMG} contamination, although motion artifacts, baseline fluctuations, and electronic noise may also contribute to signal degradation.

The objective of the denoising system is to to estimate the clean signal $x[n]$ from the noisy observation $y[n]$. 
This can be formulated as mapping the noisy sequence to an estimated clean sequence $\hat{x}[n]$ via a denoising function $\mathcal{F}(\cdot):\mathbb{R}^{N\times 1}\mapsto\mathbb{R}^{N\times 1} $. Thus,
\begin{equation}
    \hat{x}[n] = \mathcal{F}\big(y[n]\big)\Longleftrightarrow \hat{\bm{x}}=\mathcal{F}\big(\bm{y}\big)\in\mathbb{R}^{N\times 1},
\end{equation}
where $\mathcal{F}(\cdot)$ represents either a model-based denoising method or a \gls{DL} architecture.

Since the clean upper-arm \gls{ECG} signal is not directly observable, a simultaneously acquired Lead-I \gls{ECG} signal, denoted as $x_{\mathrm{ref}}[n]$, is used as a reference for performance evaluation. Although the Lead-I and upper-arm \gls{ECG} signals are acquired using different electrode configurations and therefore exhibit different amplitudes and waveform morphologies, they originate from the same cardiac activity and present temporally aligned events, such as the QRS complexes. Owing to its higher \gls{SNR} and clearer waveform morphology, $x_{\mathrm{ref}}[n]$ provides a suitable reference for assessing the preservation of clinically relevant \gls{ECG} features after denoising. 

Therefore, denoising performance can be quantitatively assessed by comparing the reconstructed signal $\hat{x}[n]$ with $x_{\mathrm{ref}}[n]$ through similarity and error-based metrics.

\section{ECG Denoising Methods}
\label{sec:denoising_methods}

This section presents the investigated ECG denoising methods, including the model-based approaches EMD and DWT, and two DL-based architectures: \glspl{SDAE} and \glspl{PINN}. Their comparison enables the evaluation of noise suppression capabilities and preservation of clinically relevant ECG features.

\subsection{Model-Based Tools} \label{sec:model_tools}

\gls{EMD} is an adaptive signal decomposition technique that decomposes the observed \gls{ECG} signal $y[n]$ into a finite set of oscillatory components, known as \glspl{IMF}, and a residual term. 
At each iteration, the local extrema of the signal are used to construct upper and lower envelopes through cubic spline interpolation. Their mean is iteratively removed until the extracted component satisfies the IMF conditions, namely: i) the numbers of extrema and zero crossings differ by at most one, and ii) the local mean is zero.
The procedure is recursively applied to the residual signal until all IMFs have been extracted.


The observed signal can therefore be represented as

\begin{equation}
    {y}[n] = \sum_{i=1}^{M} C_i[n] + r_M[n]
\end{equation}

where $C_i[n]$  denotes the $i$-th \gls{IMF}, $M$ is the total number of extracted \glspl{IMF}, and $r_M[n]$ is the final residue.


The extracted \glspl{IMF} are ordered from high- to low-frequency content. Since noise is mainly concentrated in the first modes, the denoised signal $\hat{x}[n]$ can be obtained by selectively suppressing or thresholding these components.
To identify the modes affected by noise, the logarithmic IMF energy distribution is compared with a theoretical noise energy model; the divergence point between the two distributions defines the boundary between noise- and signal-dominated IMFs \cite{Kopsinis2009EMDDenoising}.


In threshold-based schemes, denoising is performed using a scale-dependent universal threshold \cite{mohguen2019comparative}, defined as

\begin{equation}
    \tau_i = A \cdot \sqrt{\hat{E}_i \cdot 2\ln N}
\end{equation}

where $\hat{E}_i$ denotes the theoretical energy of the $i$-th \gls{IMF}, $N$ is the signal length and $A$ is an empirically tuned scaling factor. 

Several \gls{EMD}-based denoising variants were evaluated 
\begin{itemize}
    \item\textit{\gls{EMD-IT}}, 
    based on interval thresholding of IMF extrema and available in hard and soft variants;
    
    \item \textit{\gls{EMD-ITF}}, 
    employing a smooth thresholding transition 
 to reduce reconstruction discontinuities;
    
    \item \textit{\gls{EMD-Custom}}, 
    using a continuous differentiable thresholding function that combines the characteristics of hard and soft thresholding \cite{Mohguen2017EMDITF}.
\end{itemize}

Differently from \gls{EMD}, \gls{DWT} relies on a predefined basis and provides a compact time–frequency representation through multiresolution analysis. The decomposition is implemented using a pair of quadrature mirror filters (\glspl{QMF}), namely a low-pass filter g[n] and a high-pass filter h[n]. At each decomposition level L, the signal y[n] is decomposed into approximation and detail coefficients as follows \cite{seljuq2014selection}

\begin{equation}
\small
a_L[k] = \sum_n y[n]g(2k-n),
\qquad
d_L[k] = \sum_n y[n]h(2k-n)
\end{equation}

where $a_L[k]$ and $d_L[k]$ denote the approximation and detail coefficients, respectively, and the factor $2k$ accounts for the downsampling operation by a factor of two. The choice of the mother wavelet strongly influences the decomposition, as it determines the time–frequency localization properties of the transform \cite{Merry2005Wavelet}.




Through successive filtering stages, the signal is recursively decomposed into approximation coefficients, representing the low-frequency content, and detail coefficients, containing the high-frequency components. Since noise is predominantly concentrated in the detail coefficients, denoising is typically performed by thresholding these coefficients while preserving the approximation coefficients. The denoised signal is finally reconstructed using the inverse \gls{DWT}.

In this work, soft thresholding was adopted due to its ability to better preserve the morphology of the \gls{ECG} while reducing reconstruction artifacts. The thresholding operation is defined as \cite{Shemi2016ECGDenoising}

\begin{equation}
\small
\tilde{d}_{L}[k] =
\begin{cases}
\mathrm{sign}(d_L[k])\left(|d_L[k]|-\lambda_L\right), & |d_L[k]| > \lambda_L, \\
0, & |d_L[k]| \leq \lambda_L.
\end{cases}
\end{equation}

where $\tilde{d}_{L}[k]$ denotes the threshold detail coefficient and $\lambda_L$ is the threshold value applied at the decomposition level $L$ \cite{Verma2012WaveletThresholding}.


\subsection{Deep Learning-Based Tools}
\label{sec:dl_tools}
The denosing function $\mathcal{F}$ can be learned by a proper \gls{DL} tool by providing some training data. The \gls{DAE} is a traditional approach for denoising purposes and it consists of two parameterized functions:
    \begin{itemize}
        \item  \textit{Encoder} $\bm{f}(\cdot|\bm{\Theta}_{\text{e}}):\mathbb{R}^{N\times 1}\mapsto\mathbb{R}^{M\times 1} $ that maps the noisy input $\bm{y}\in\mathbb{R}^{N\times 1}$ to a latent representation $\bm{z}\in\mathbb{R}^{M\times 1}$, i.e., 
\begin{equation}\label{EQ:  latent representation}
\begin{split}
            \bm{z} 
            &= \bm{f}(\bm{y}|\bm{\Theta}_{\text{e}})\\
            & = \bm{\sigma}_{\text{e}} (\bm{W}_{\text{e}} \bm{y} + \bm{b}_{\text{e}})
\end{split}
\end{equation}
        
        and

        \item \textit{Decoder} $\bm{g}(\cdot|\bm{\Theta}_{\text{d}}):\mathbb{R}^{M\times 1}\mapsto\mathbb{R}^{N\times 1}$ that reconstructs the clean signal $\bm{\hat{x}}\in\mathbb{R}^{N\times 1}$ from the latent space $\bm{z}\in\mathbb{R}^{M\times 1}$, i.e.,
\begin{equation}
\begin{split}
            \bm{\hat{x}} &= \bm{g}(\bm{z}|\bm{\Theta}_{\text{d}})\\
            & = \bm{\sigma}_{\text{d}} (\bm{W}_{\text{d}} \bm{z} + \bm{b}_{\text{d}} ),
        \end{split}
\end{equation}
    \end{itemize}
where $\bm{\sigma}_{\text{e}} (\cdot)$ and $\bm{\sigma}_{\text{d}} (\cdot)$ denote element-wise nonlinear activation functions, $\bm{W}_{\text{e}} \in \mathbb{R}^{M \times N}$ and $\bm{W}_{\text{d}} \in \mathbb{R}^{N \times M}$ are the weight matrices, with $M\in\mathbb{N}^{1\times 1}$ being the dimension of the latent space  or the number of neurons (nodes) in the hidden layer, $\bm{b}_{\text{e}} \in \mathbb{R}^{M \times 1}$ and $\bm{b}_{\text{d}} \in \mathbb{R}^{N \times 1}$ are the bias vectors,
and $\bm{\Theta}_{\text{e}}\in \mathbb{R}^{M \times (N+1)}$ and $\bm{\Theta}_{\text{s}}\in \mathbb{R}^{ (N+1)\times M}$ are the sets of learnable parameters for the encoder and decoder, respectively, given by
\begin{subequations}
\begin{align}
\bm{\Theta}_{\text{e}} &= 
\begin{bmatrix} \bm{W}_{\text{e}} & \bm{b}_{\text{e}} \end{bmatrix},\\
\bm{\Theta}_{\text{s}} &= 
\begin{bmatrix} \bm{W}_{\text{d}} & \bm{b}_{\text{d}}  \end{bmatrix}.
\end{align}
\end{subequations}
\tikzset{
    ecgline/.style={thick, red!80!black, line join=round},
    axisline/.style={->, thick, gray},
    labeltext/.style={font=\footnotesize\sffamily, align=center},
    mainlabel/.style={font=\small\sffamily\bfseries},
    annotationbox/.style={
        draw=blue!50!black,
        fill=blue!5,
        rounded corners,
        align=left,
        font=\footnotesize\sffamily,
        inner sep=6pt,
        shadow={4pt}{-4pt}{0pt}{black!20}
    },
    signalbox/.style={
        draw=orange!70!black,
        fill=orange!10,
        rounded corners,
        align=left,
        font=\footnotesize\sffamily,
        inner sep=6pt
    }
}

\tikzset{
    block/.style={
        draw=blue!80!black, 
        fill=blue!10, 
        rounded corners, 
        thick, 
        align=center, 
        font=\sffamily\footnotesize
    },
    noiseblock/.style={
        block,
        draw=orange!80!black,
        fill=orange!20
    },
    signalbox/.style={
        draw=gray!60!black, 
        fill=white, 
        thick, 
        minimum width=2.5cm, 
        minimum height=1.8cm,
        align=center,
        font=\sffamily\footnotesize
    },
    latentbox/.style={
        signalbox,
        minimum width=1.5cm, 
        minimum height=2.5cm,
        fill=gray!10
    },
    connect/.style={
        ->, 
        >=Latex, 
        very thick, 
        blue!80!black
    },
    lossconnect/.style={
        <->,
        >=Latex,
        very thick,
        dashed,
        red!80!black
    },
    cleanwave/.style={blue!80!black, thick, smooth},
    noisywave/.style={orange!90!black, thin, decorate, decoration={random steps, segment length=1pt, amplitude=2pt}}
}

\tikzset{
    device block/.style={
        rectangle, 
        draw=blue!80!black, 
        fill=blue!10, 
        rounded corners, 
        minimum height=3cm, 
        minimum width=2.5cm, 
        align=center, 
        font=\sffamily\small
    },
    sensor/.style={
        circle, 
        draw=gray!80!black, 
        fill=gray!40, 
        inner sep=3pt,
        label={[font=\footnotesize]below:Electrode}
    },
    wire/.style={
        draw, 
        very thick, 
        gray!70!black, 
        rounded corners=5pt
    },
    flow arrow/.style={
        ->, 
        >=Latex, 
        ultra thick, 
        blue!80!black
    },
    waveform line/.style={
        red!80!black,
        thick,
        line join=round,
        decorate, decoration={random steps, segment length=1pt, amplitude=0.2pt}
    }
}

The learnable parameters $\bm{\Theta}_{\text{e}}$ and $\bm{\Theta}_{\text{d}}$ are trained by minimizing the reconstruction error, which is typically \gls{MSE}, between  the output and the original clean input, i.e., 
\begin{equation}\label{EQ: the reconstruction error in mean squared error}
\begin{split}
\begin{bmatrix}
\bm{\hat{\Theta}}_{\text{e}} & \bm{\hat{\Theta}}_{\text{d}}
\end{bmatrix}
&=\underset{\bm{\Theta}_{\text{e}}, \bm{\Theta}_{\text{d}}}{\arg\min} f_{\text{MSE}}(\bm{\Theta}_{\text{e}}, \bm{\Theta}_{\text{d}}) \\
&=\underset{\bm{\Theta}_{\text{e}}, \bm{\Theta}_{\text{d}}}{\arg\min}    \sum_{b=1}^{B} \| \bm{\hat{x}}[b] - \bm{x}[b]  \|_2^2,
\end{split}
\end{equation} 
 where $B$ represents the  batch size  (number of samples per update),
$b \in\{1,2,\ldots,B\}$ is the sample index within the training batch, and $\| \cdot \|_2$ is the Euclidean norm.

For the comparison, the following methods are considered.

\subsubsection{Stacked Denoising Autoencoder (SDAE)}
Instead of trying to learn everything in a single shallow step, a stacked architecture breaks the learning process into manageable unsupervised steps ultimately building a deep model capable of performing exceptionally well on complex recognition tasks.
The \gls{SDAE} is a DL model used in machine learning for unsupervised learning and feature extraction. It consists of multiple \glspl{DAE} stacked together, which learn to reconstruct clean data from a noisy or corrupted input. The input signal $\bm{y}$ is passed through a sequence of $L$ hidden layers. As the signal passes through each layer, the network learns increasingly complex, higher-level, and non-linear representations of the data. During training, the network  output ($\bm{\hat{x}}$) is compared to the known ground-truth clean signal ($\bm{x}$), and the errors are backpropagated to adjust the weights of the hidden layers. The network is trained to ignore the noise and recover the original uncorrupted signal.

\subsubsection{Physics-Informed Neural Network (PINN)}

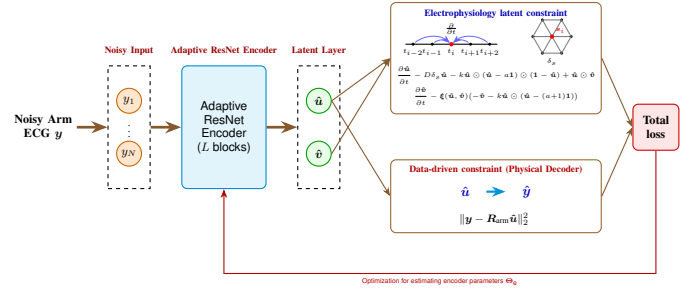
\begin{figure}[t]
\centering
\resizebox{\columnwidth}{!}{%
\begin{tikzpicture}[
    >=Stealth,
    node distance=0.8cm,
    font=\sffamily\small,
    dashedbox/.style={
        draw=black, dashed, thick,
        minimum width=1.1cm, minimum height=3.2cm, align=center
    },
    circleNode/.style={
        draw=orange!80!black, fill=orange!20,
        circle, thick, minimum size=0.65cm, inner sep=1pt
    },
    resblock/.style={
        draw=cyan!70!blue, fill=cyan!10, thick, rounded corners,
        minimum width=2.2cm, minimum height=3.2cm, align=center
    },
    constraintbox/.style={
        draw=brown!80!black, thick, rounded corners,
        minimum width=5.5cm, minimum height=2.9cm,
        align=center, fill=white
    },
    lossbox/.style={
        draw=red!80!black, fill=red!10, thick, rounded corners,
        minimum width=1.2cm, minimum height=1.2cm,
        align=center, font=\bfseries\small
    }
]

\node (heart) [font=\bfseries\small, align=center] {Noisy Arm\\ECG $\bm{y}$};

\node[dashedbox, right=0.8cm of heart] (inputlayer) {};
\node[circleNode, above=0.35cm of inputlayer.center] (y_top)  {$y_1$};
\node[font=\footnotesize] at (inputlayer.center) {$\vdots$};
\node[circleNode, below=0.35cm of inputlayer.center] (y_bot)  {$y_N$};
\node[above=0.15cm of inputlayer,
      font=\bfseries\scriptsize\color{red!60!black}] {Noisy Input};

\node[resblock, right=0.8cm of inputlayer] (resnet)
    {Adaptive\\ResNet\\Encoder\\($L$ blocks)};
\node[above=0.15cm of resnet,
      font=\bfseries\scriptsize\color{red!60!black}] {Adaptive ResNet Encoder};

\node[dashedbox, right=0.8cm of resnet] (outputlayer) {};
\node[circleNode, above=0.35cm of outputlayer.center,
      draw=green!60!black, fill=green!10] (u_out) {$\bm{\hat{u}}$};
\node[circleNode, below=0.35cm of outputlayer.center,
      draw=green!60!black, fill=green!10] (v_out) {$\bm{\hat{v}}$};
\node[above=0.15cm of outputlayer,
      font=\bfseries\scriptsize\color{red!60!black}] {Latent Layer};

\node[constraintbox, right=1.3cm of outputlayer, yshift=1.8cm] (ep) {};
\node[below=0.08cm of ep.north,
      font=\bfseries\scriptsize\color{blue!70!black}]
    {Electrophysiology latent constraint};
\node[above=0.12cm of ep.south, font=\tiny, align=center] {
    $\dfrac{\partial \bm{\hat{u}}}{\partial t}
     - D\delta_{s}\bm{\hat{u}}
     - k\bm{\hat{u}}\odot(\bm{\hat{u}}-a\bm{1})\odot(\bm{1}-\bm{\hat{u}})
     + \bm{\hat{u}}\odot\bm{\hat{v}}$\\[1.0ex]
    $\dfrac{\partial \bm{\hat{v}}}{\partial t}
     - \bm{\xi}(\bm{\hat{u}},\bm{\hat{v}})
       \bigl(-\bm{\hat{v}} - k\bm{\hat{u}}\odot(\bm{\hat{u}}-(a{+}1)\bm{1})\bigr)$
};

\begin{scope}[shift={(ep.north west)}, xshift=1.6cm, yshift=-1.1cm]
    \draw[thick] (-1.2,0) -- (1.2,0);
    
    \foreach \x/\lbl in {-1.0/{t_{i-2}}, -0.5/{t_{i-1}}, 0/{t_i}, 0.5/{t_{i+1}}, 1.0/{t_{i+2}}} {
        \fill[black] (\x,0) circle (1.1pt);
        \node[below, font=\tiny] at (\x,0) {$\lbl$};
    }
    
    \fill[red] (0,0) circle (1.6pt)
        node[above=0.15cm, font=\tiny, text=black] {$\tfrac{\partial}{\partial t}$};
        
    \draw[->, blue!60, shorten >=2pt, shorten <=2pt]
        (-1.0,0) to[bend left=40] (0,0);
    \draw[->, blue!60, shorten >=2pt, shorten <=2pt]
        (-0.5,0) to[bend left=40] (0,0);
        
    \draw[->, blue!60, shorten >=2pt, shorten <=2pt]
        ( 1.0,0) to[bend right=40] (0,0);
    \draw[->, blue!60, shorten >=2pt, shorten <=2pt]
        ( 0.5,0) to[bend right=40] (0,0);
\end{scope}

\begin{scope}[shift={(ep.north east)}, xshift=-1.3cm, yshift=-0.9cm]
    \coordinate (c) at (0,0);
    \foreach \a in {0, 60, 120, 180, 240, 300} {
        \coordinate (p\a) at (\a:0.5);
        \draw[gray, thick] (c) -- (p\a);
        \fill[black] (p\a) circle (1.1pt);
    }
    \draw[gray, thick] (p0)--(p60)--(p120)--(p180)--(p240)--(p300)--cycle;
    \fill[red] (c) circle (1.6pt)
        node[above right=-1pt, font=\tiny, text=red] {$\bm{s}_i$};
    \node[below=0.45cm, font=\tiny] at (0,0) {$\delta_{s}$};
\end{scope}

\node[constraintbox, minimum height=1.9cm,
      right=1.3cm of outputlayer, yshift=-1.8cm] (data) {};

\node[below=0.05cm of data.north,
      font=\bfseries\scriptsize\color{red!70!black}]
    {Data-driven constraint (Physical Decoder)};

\node[above=0.08cm of data.south, font=\footnotesize, align=center] {
    $\|\bm{y} - \bm{R}_{\text{arm}}\bm{\hat{u}}\|^{2}_{2}$
};

\begin{scope}[shift={(data.center)}, yshift=0.12cm]
    \node[font=\normalsize, color=blue!80!black] at (-0.8, 0) {$\bm{\hat{u}}$};
    \draw[->, ultra thick, cyan!80!blue] (-0.25,0) -- (0.25,0);
    \node[font=\normalsize, color=blue!80!black] at ( 0.8, 0) {$\bm{\hat{y}}$};
\end{scope}

\path (ep.east) -- (data.east) coordinate[midway] (midRight);
\node[lossbox, right=0.8cm of midRight] (loss) {Total\\loss};

\draw[->, line width=2pt, brown!70!black] (heart)      -- (inputlayer);
\draw[->, line width=2pt, brown!70!black] (inputlayer) -- (resnet);
\draw[->, line width=2pt, brown!70!black] (resnet)     -- (outputlayer);

\draw[->, thick, brown!70!black] (u_out.east) -- (ep.west);
\draw[->, thick, brown!70!black] (v_out.east) -- (ep.west);
\draw[->, thick, brown!70!black] (u_out.east) -- (data.west);

\draw[->, thick, brown!70!black] (ep.east)   -- (loss.west);
\draw[->, thick, brown!70!black] (data.east) -- (loss.west);

\coordinate (feedDown) at ($(loss.south)+(0,-3.2cm)$);
\draw[->, thick, red!70!black]
    (loss.south)
    -- (feedDown)
    -- (resnet.south |- feedDown)
    -- (resnet.south);
\node[below, font=\tiny\sffamily\color{red!70!black}]
    at ($(feedDown)!0.5!(resnet.south |- feedDown)$)
    {Optimization for estimating encoder parameters $\bm{\Theta}_{\text{e}}$};

\end{tikzpicture}%
}
\caption{Physics-informed adaptive residual denoising framework for arm ECG processing.}
\label{fig:eand-arn_framework}
\end{figure}

The architecture of the   adaptive residual physics-informed denoising  framework using the Aliev-Panfilov electrophysiology model, automatic numerical differentiation, and an adaptive residual network is shown in Fig. \ref{fig:eand-arn_framework}.
The encoder $\bm{f}(\cdot|\bm{\Theta}_{\text{e}})$ takes a corrupted or noisy version of the arm ECG measurements at a specific time instance $t$. 
An adaptive residual network structure is utilized inside the encoder to avoid training instability, low derivative variance, or ineffective scaling when the parameter layers are made deeper. 
The relation between the input and output of the $l$-th adaptive residual block (Adaptive ResBlock) within the encoder can be expressed as \cite[eq. (12)]{2025:Zhu_et_al}
\begin{subequations}
\begin{align}
    \bm{\hat{h}}_1[l] &= \bm{\sigma} ( \bm{W}_1[l]\bm{\hat{x}}[l] + \bm{b}_1[l] ), \\
    \bm{\hat{h}}_2[l] &= \bm{\sigma} ( \bm{W}_2[l]\bm{\hat{h}}_1[l] + \bm{b}_2[l] ), \\
    \bm{\hat{x}}[l+1] &= (1 - \alpha[l])\bm{\hat{h}}_2[l] + \alpha[l]\bm{\hat{x}}[l],
\end{align}
\end{subequations}
where $\bm{\hat{x}}[1] = \bm{y}$ represents the initial noisy input vector to the first block, $\bm{\hat{h}}_1[l]$ and $\bm{\hat{h}}_2[l]$ are the outputs from the hidden layers within the $l$-th block, $\bm{W}_1[l]$ and $\bm{W}_2[l]$ are the weight matrices, $\bm{b}_1[l]$ and $\bm{b}_2[l]$ are the bias vectors, and $\alpha[l]$ is a trainable parameter that controls the proportion of knowledge bypassed via the identity skip connection.

The adaptive residual encoder compresses the noisy inputs into a physically meaningful latent bottleneck space. This latent space yields two primary hidden spatial fields modeled simultaneously across all $N_{\mathrm{h}}$ virtual nodes of the 3D heart geometry: the estimated true heart surface potential $\bm{\hat{u}}\in\mathbb{R}^{N N_{\mathrm{h}} \times 1}$ and the corresponding recovery current $\bm{\hat{v}}\in\mathbb{R}^{N N_{\mathrm{h}} \times 1}$, which regulates local cell depolarization. The PINN evaluates  a dual-objective loss function split into two constraint branches: an electrophysiology (EP) latent constraint and a data-driven (DD) reconstruction constraint.

The EP branch acts directly as a biophysical regularizer on the encoder's latent space, ensuring the extracted features adhere strictly to known physical laws and preventing the autoencoder from passing physically impossible noise or artifacts into the bottleneck. The EP constraint embeds the partial differential equations (PDEs) of the Aliev-Panfilov model, which describes cardiac excitation waves. This framework applies numerical differentiation with spatiotemporal local support directly over consecutive latent states across time. Temporal support $\frac{\partial}{\partial t}$ evaluates time derivatives at time $t_{i}$ by checking a localized neighborhood of preceding ($t_{i-2}$, $t_{i-1}$) and succeeding ($t_{i+1}$, $t_{i+2}$) latent state frames. Spatial support ($\delta_{s}$) evaluates the spatial Laplacian on the irregular, complex 3D triangular heart surface by factoring in the geometric connections of neighboring nodes ($s_{j}$) surrounding the central node ($s_{i}$). These PDEs involve cellular excitation parameter constants $k$ and $a$, an all-ones vector $\bm{1}$ for element-wise subtraction compatibility to the Hadamard product $\odot$, and $\bm{\xi}(\bm{\hat{u}},\bm{\hat{v}})\in\mathbb{R}^{N N_{\mathrm{h}} \times 1}$ that represents the adaptive temporal rate function computed across the local latent temporal configuration at $t_i$.

The DD branch  enforces the clean signal reconstruction criteria. The decoder utilizes a pre-computed linear forward heart-to-arm transfer matrix $\bm{R}_{\text{arm}}\in\mathbb{R}^{N \times NN_{\mathrm{h}} }$ to project the clean latent heart potentials $\bm{\hat{u}}$ back out to the arm surface measurement domain, yielding   $\bm{\hat{y}}= \bm{R}_{\text{arm}}\bm{\hat{u}}$. The data-driven penalty $\|\bm{y}-\bm{R}_{\text{arm}}\bm{\hat{u}}\|^{2}_{2}$ calculates the  reconstruction error between the input arm ECG signal $\bm{y}\in\mathbb{R}^{N\times 1}$ and the network's projected  reconstruction $\bm{\hat{y}}$.

At the bottom of Fig. \ref{fig:eand-arn_framework}, the reconstruction errors from the arm domain and the biophysical deviations from the latent EP domain are backpropagated to iteratively estimate and update   weights, biases, and adaptive skip coefficients.

\section{Simulation Results}
\label{sec:results}

\subsection{Experimental Setup for ECG Acquisition}
\label{Experimental_Setup}

\begin{figure}[t]
    \centering
    \includegraphics[width=1 \linewidth]{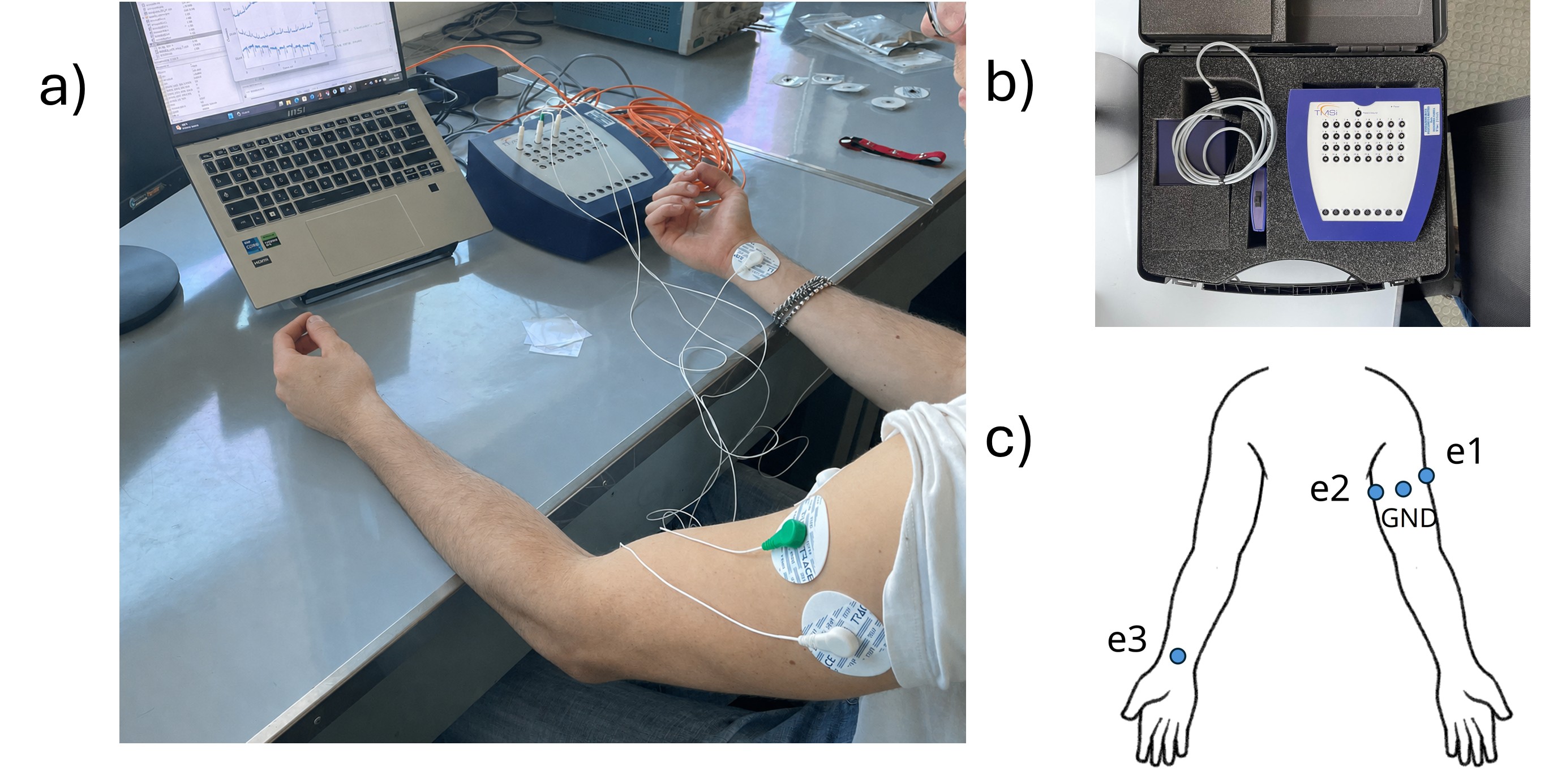}
    \caption{a) Experimental setup; b) REFA (TMSi) multichannel amplifier; c) Electrode placement for simultaneous acquisition of the upper-arm ECG signal $y_{\mathrm{UA}}$ and the reference Lead-I ECG signal $x_{\mathrm{ref}}$. The ground electrode is marked as GND. }
    \label{ElePos}
\end{figure}

Fig.~\ref{ElePos} presents the experimental setup adopted for the acquisition of upper-arm and reference \gls{ECG} signals. An overview of the measurement system is shown in Fig.~\ref{ElePos}(a), while the REFA multichannel amplifier (Twente Medical Systems
International, Oldenzaal, The Netherlands) and the electrode placement are reported in Fig.~\ref{ElePos}(b) and Fig.~\ref{ElePos}(c), respectively.

The electrode placement was selected to replicate the sensing geometry proposed in~\cite{11059393} for continuous upper-arm \gls{ECG} monitoring using a wearable armband device. Two electrodes were positioned on the left upper arm to acquire the upper-arm \gls{ECG} signal, while an additional electrode placed on the right wrist was used to obtain the Lead-I reference signal. A fourth electrode connected to the patient ground input was positioned between the two arm electrodes to stabilize the common-mode potential.

The REFA amplifier acquires unipolar measurements referenced to the average value of all active input channels. 
Let $e_1$, $e_2$, and $e_3$ denote the potentials measured by the three acquisition electrodes. The upper-arm \gls{ECG} signal, denoted as $y_{\mathrm{UA}}$, is obtained as the differential measurement between $e_1$ and $e_2$
\begin{equation}
\small
\label{eq_REFA}
y_{\mathrm{UA}}
=
e_1-e_2
=
\left(e_1-\frac{1}{N}\sum_{i=1}^{N}e_i\right)
-
\left(e_2-\frac{1}{N}\sum_{i=1}^{N}e_i\right)
\end{equation}


where $N$ is the number of active channels ($N=3$ in the considered configuration). Similarly, the reference Lead-I \gls{ECG} signal, denoted as $x_{\mathrm{ref}}$, is obtained as the difference between  $e_1$ and $e_3$.
This configuration enables the simultaneous acquisition of the upper-arm \gls{ECG} signal and the Lead-I reference \gls{ECG} signal, allowing a direct comparison between the denoised upper-arm \gls{ECG} and the reference trace.

\begin{table}[t]
\centering
\caption{Experimental acquisition scenarios.}
\label{scenarios}
\begin{tabular}{c|c|c}
\hline
Scenario & Upper-arm condition & Acquired signals \\
\hline
$S_{1}$ & Relaxed arm & $y_{r}$, $x_{\mathrm{ref}}$ \\

$S_{2}$ & Contracted arm muscles & $y_{c}$, $x_{\mathrm{ref}}$\\
\hline
\end{tabular}
\end{table} 

To evaluate the robustness of the proposed denoising methods, two acquisition scenarios were considered during the experimental campaign, corresponding to a moderate and severe noise condition, respectively. In both cases, the upper-arm \gls{ECG} and the reference Lead-I \gls{ECG} were acquired simultaneously.
Tab.~\ref{scenarios} summarizes the two experimental scenarios. In the first scenario ($S_{1}$), the subject maintained the arm in relaxed state in order to acquire an upper-arm \gls{ECG} signal affected by limited muscular interference ($y_{r}$). In the second scenario ($S_{2}$), the subject intentionally contracted the arm muscles during the acquisition, resulting in the upper-arm \gls{ECG} signal $y_{c}$, affected by increased \gls{EMG} contamination. 


For \gls{DL}-based methods, additional clean \gls{ECG} recordings collected under controlled conditions, denoted as $x_{\mathrm{clean}}$, are used for training.

\subsection{Performance Metrics}
\label{metrics}

To compare \gls{DL}-based and model based denoising approaches, the following performance metrics are considered.
\subsubsection{Peak-to-peak signal-to-noise ratio}
The quality of the reconstructed \gls{ECG} signal is measured via \gls{PPSNR}, that is defined as the ratio of the peak-to-peak amplitude of the \gls{ECG} signal to the peak-to-peak amplitude of the background noise \cite{2017:Escalona_et_al},  i.e.,
\begin{equation}\label{EQ: SNR definition}
\begin{split}
\mathrm{PPSNR}(\bm{x}) &= \frac{\mathrm{PPQRS}}{\mathrm{PPN}},
\end{split}
\end{equation}
where  $\mathrm{PPQRS}$ denotes the peak-to-peak amplitude of the QRS complex measured within a $120$ \SI{}{\milli\second} time window centered on the R-wave     and
$\mathrm{PPN}$ denotes the peak-to-peak amplitude of the noise measured within a $40$ \SI{}{\milli\second} time window positioned in the T-P interval (the isoelectric segment between the end of the T-wave and the start of the next P-wave).

\subsubsection{Pearson correlation coefficient}
The \gls{PCC} ($c_{\text{p}} $)  is often adopted to quantitatively assess the morphological preservation and structural similarity between the denoised \gls{ECG} output and the ground truth signal, given by \cite[eq. (8)]{2008:Benesty_et_al}
        \begin{equation}\label{EQ: pearson correlation coefficient}
        \begin{split}
        c_{\text{p}} (\bm{\hat{x}}|\bm{x}) &= \frac{\sum\limits^{N}_{n=1}(x[n] - \bar{x})(\hat{x}[n] - \bar{\hat{x}})}{\sqrt{\sum\limits^{N}_{n=1}(x[n] - \bar{x})^{2}} \sqrt{\sum\limits^{N}_{n=1}(\hat{x}[n] - \bar{\hat{x}})^{2}}} \\
        &= \frac{(\bm{x} - \bar{x}\bm{1}_{N})^{\T}(\bm{\hat{x}} - \bar{\hat{x}}\bm{1}_{N})}{\|\bm{x} - \bar{x}\bm{1}_{N}\|_{2} \|\bm{\hat{x}} - \bar{\hat{x}}\bm{1}_{N}\|_{2}},
        \end{split}
        \end{equation}
        where   $\bar{x}$ and $\bar{\hat{x}}$ denote the mean values of the true signal $\bm{x}$ and the estimated signal $\bm{\hat{x}}$, respectively, $\bm{1}_{N}\in\mathbb{R}^{N\times 1}$ is an all-ones vector of length $N$,  $\cdot^{\T}$ is the transpose of a vector or matrix, and $\|\cdot\|_{2}$ is the $\ell_{2}$ norm.  
\subsubsection{Root mean squared error}
The \gls{RMSE} is a standard metric in \gls{ECG} signal processing to quantify the absolute point-wise amplitude between the reconstructed signal and the ground truth, given by 
\begin{equation}\label{EQ: root-mean-squared error (RMSE)}
\begin{split}
\mathrm{RMSE}(\bm{\hat{x}}|\bm{x})
&=\sqrt{
\frac{1}{N}
\sum\limits^{N}_{n=1}
(\hat{x}_{n}-x_{n})^{2}
} \\
&=\sqrt{
\frac{1}{N}
\|\bm{\hat{x}}-\bm{x}\|^{2}_{2}
} .
\end{split}
\end{equation}

\begin{figure*}[htbp]
    \centering
    \includegraphics[width=\textwidth]{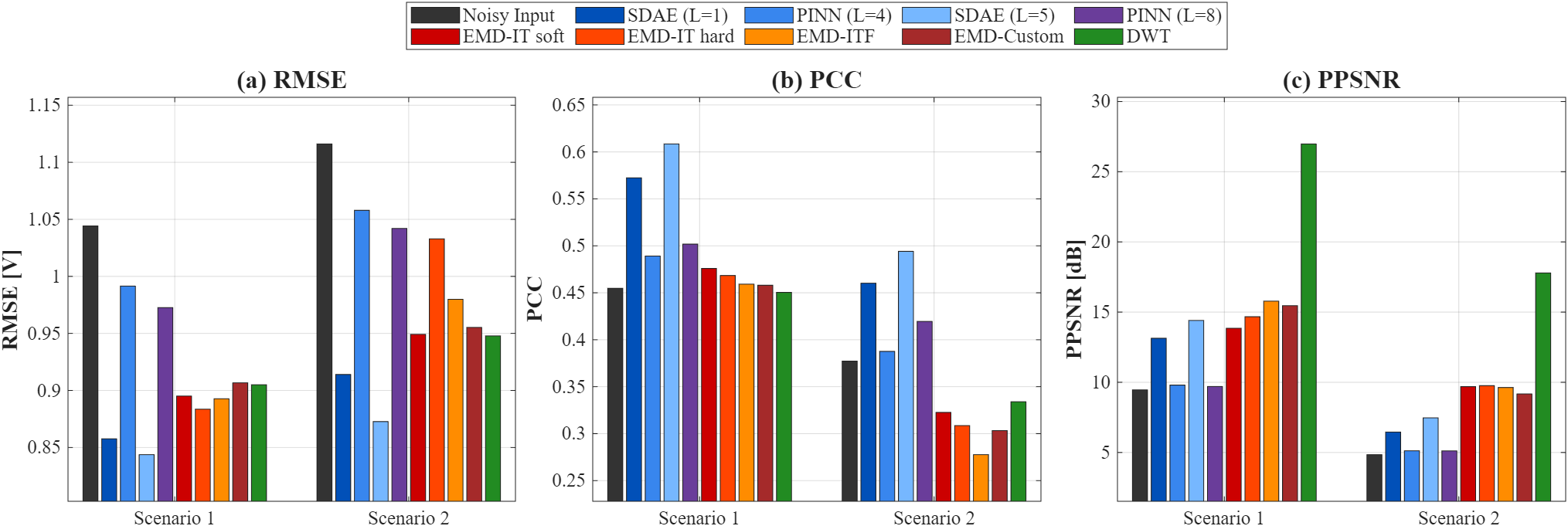}
    \caption{Performance comparison of the evaluated denoising methods across Scenario 1 and Scenario 2. The grouped bar charts report the achieved root mean square error (RMSE), Pearson correlation coefficient (PCC), and peak-to-peak signal-to-noise ratio (PPSNR) for both the DL architectures and the traditional model-based techniques, highlighting their relative effectiveness in signal reconstruction under the different analyzed conditions.}
    \label{fig:metrics_comparison}
\end{figure*}

\subsection{Discussion}

Figure \ref{fig:metrics_comparison} shows a comparative analysis of model-based and \gls{DL} denoising techniques and it reveals distinct performance trade-offs. Across all evaluated metrics, the introduction of severe \gls{EMG} interference due to voluntary muscle contraction in Scenario 2 systematically degrades denoising performance compared to the relaxed arm condition in Scenario 1.

In terms of point-wise signal reconstruction, measured by the \gls{RMSE}, \gls{DL} methods generally demonstrate superior accuracy. Among the model-based techniques, \gls{EMD-IT} hard performs best in Scenario 1 (0.8836), while the \gls{DWT} is the most effective in Scenario 2 (0.9478). Among the \gls{DL} models, the \gls{SDAE} with 5 hidden layers (SDAE L=5) achieves the best performance, yielding the lowest \gls{RMSE} in both scenarios (0.8437 in S1 and 0.8727 in S2). Overall, the \gls{SDAE} (L=5) proves to be the absolute best method for minimizing \gls{RMSE} across both noise conditions.

Morphological preservation, evaluated via the \gls{PCC}, exhibits a similar trend. The \gls{SDAE} (L=5) is again the highest-performing \gls{DL} method and the absolute best overall, reaching a \gls{PCC} of 0.6085 in S1 and 0.4941 in S2, indicating a strong ability to track the underlying cardiac waveforms. Conversely, traditional methods struggle to match this fidelity. Among model-based tools, \gls{EMD-IT} soft yields the highest \gls{PCC} in S1 (0.4759), whereas \gls{DWT} leads in S2 (0.3340). Notably, under severe noise (S2), \gls{EMD} variants show a significant degradation in morphological tracking—often falling below the \gls{PCC} of the raw noisy input—leaving \gls{DWT} as the most robust traditional alternative, though still inferior to the \gls{SDAE} architecture.

However, the evaluation of overall noise suppression via the \gls{PPSNR} reveals a completely different hierarchy. In this domain, the \gls{DWT} exhibits undisputed dominance. \gls{DWT} not only represents the best model-based approach but also the absolute best method overall for noise attenuation, achieving an impressive \gls{PPSNR} of 26.98 dB in S1 and 17.79 dB in S2. While \gls{SDAE} (L=5) remains the best among \gls{DL}-based methods (14.41 dB in S1 and 7.47 dB in S2), it falls significantly short of \gls{DWT}'s filtering capabilities. Furthermore, in terms of \gls{PPSNR}, the \gls{SDAE} is even outperformed by several \gls{EMD} variants, such as \gls{EMD-ITF} in S1 (15.78 dB) and \gls{EMD-IT} hard in S2 (9.76 dB).

This clear dichotomy highlights a fundamental trade-off in upper-arm \gls{ECG} denoising: while \gls{DL} networks (specifically \gls{SDAE}) excel at mapping the underlying signal morphology and reducing point-wise errors (yielding the best \gls{PCC} and \gls{RMSE}), the highly ill-posed nature of the \gls{ECG} inverse problem limits their ability to suppress background noise entirely. On the other hand, traditional model-based methods like \gls{DWT} offer superior noise filtering and background suppression (yielding the highest \gls{SNR}) but compromise the exact temporal morphology and absolute point-wise accuracy of the cardiac features compared to \gls{DL} approaches.

\section{Conclusion and Future Work}
\label{sec:conclusion}

This paper presented a comparative evaluation of model-based and \gls{DL}-based \gls{ECG} denoising techniques for upper-arm wearable recordings,
motivated by the demand for reliable cardiac monitoring in \gls{WSN} deployments and
space environments. The methods assessed include three \gls{EMD}-based variants
(EMD-IT, EMD-ITF, EMD-Custom), a \gls{DWT}-based approach, and two \gls{DL}
architectures (\gls{SDAE} and \gls{PINN}) evaluated on real
acquisitions under relaxed~(S1) and contracted-muscle~(S2) conditions.

The results reveal a consistent performance trade-off between the two families
of methods. In terms of point-wise accuracy~(\gls{RMSE}) and morphological
fidelity~(\gls{PCC}), the \gls{SDAE} with five hidden layers achieves the best overall
performance. In contrast, the \gls{DWT}-based approach
demonstrates unambiguous superiority in noise suppression, substantially exceeding
all competing methods. This dichotomy reflects the ill-posed nature of the \gls{ECG}
inverse problem: \gls{DL} networks excel at morphological mapping but struggle to
fully suppress residual background noise, while \gls{DWT} achieves strong noise
attenuation at the cost of reduced point-wise accuracy. All methods exhibit
marked degradation under severe \gls{EMG} contamination~(S2), due to the spectral
overlap between muscular interference and ECG components.

Future work will focus on hybrid architectures combining the noise-suppression
of \gls{DWT} or \gls{EMD} with the morphological learning of \gls{DL} models, as well as
lightweight implementations suitable for real-time inference on embedded \gls{WSN}
nodes, targeting end-to-end wearable systems for space health monitoring.

\ifCLASSOPTIONcaptionsoff
  \newpage
\fi

\balance
\bibliographystyle{IEEEtran}
\bibliography{references}

\end{document}